\documentclass{ws-procs9x6}
\usepackage[latin1]{inputenc}
\usepackage{wrapfig,rotating}
\usepackage{amssymb,amsmath,array}

\begin{document}
\title{Generalized Parton Distributions from Hadronic Observables}

%***********************************************************************
% AUTHORS INFORMATION AREA
%***********************************************************************
\author{S. Ahmad, H. Honkanen, S. Liuti}

\address{University of Virginia - Physics Department \\
382, McCormick Rd., Charlottesville, Virginia 22904 - USA}
\author{S.K. Taneja}

\address{Ecole Polytechnique, \\ CPHT, F91128 Palaiseau Cedex, France}

%***********************************************************************
% END OF AUTHORS INFORMATION AREA
%***********************************************************************

\begin{abstract}
Following a previous detailed study of unpolarized generalized parton distribution
functions in the non-singlet sector, and at zero values
of the skewness variable, $\zeta$, we propose a physically motivated 
parametrization that is valid at $\zeta \neq 0$. 
Our method makes use of information from the nucleon form factor data, 
from deep inelastuc scattering parton distribution functions, and from lattice 
results on the Mellin moments of generalized parton distributions. 
It provides, therefore, a step towards a model independent extraction 
of generalized distributions from the data,
alternative to the mathematical ansatz of 
double distributions. Comparisons with recent experimental data on the proton 
are shown. 
\end{abstract}

%***************************************************************************
\bodymatter
\section{Introduction}
Generalized Parton Distributions (GPDs) parametrize the soft  
contributions in a variety of hard exclusive 
processes, from Deeply Virtual Compton Scattering (DVCS)
to hard exclusive meson production (see \cite{Die_rev,BelRad} for reviews).
The feasibility of DVCS-type experiments using a deep inelastic 
probe with virtuality, $Q^2$, while detecting a momentum
transfer, $t$, between the initial and final proton, 
allows one to address a vast, previously inaccessible 
phenomenology. In particular, one can access parton densities
in impact parameter space \cite{Bur}, and envisage extracting 
the orbital angular momentum of partons in both nucleons and nuclei \cite{Ji1}.

At present, a central issue is the definition of a quantitative, reliable 
approach beyond the construction of GPDs from specific models and/or 
particular 
limiting cases, that can incorporate new incoming experimental data 
in a variety of ranges of $Q^2$, and of the longitudinal (along the lightcone) 
and transverse components of the four-momentum transfer 
between the incoming
and outgoing protons, $\zeta$ and $\Delta_\perp$, respectively (see also discussion
in \cite{KumMue}). 
The matching between measured quantities and Perturbative QCD (PQCD) based predictions   
for DVCS should proceed, owing to s pecific factorization theorems, 
similarly to the extraction of Parton Distribution Functions (PDFs) from deep 
inelastic scattering. 
A few important caveats are however present since
GPDs describe {\em amplitudes} and are therefore more elusive observables.
The comparison between experiment and the formulation of parametrizations 
necessarily encompasses other strategies using additional constraints, 
since a direct comparison with the data cannot be performed.     
Experiments providing sufficiently accurate data to constrain the shape of 
GPDs have just begun \cite{hallap,hallan}. 

In Refs.\cite{AHLT1,AHLT2} we proposed a strategy using a combination of experimental 
data on nucleon form factors, PDFs, and lattice calculations of Mellin moments 
with $n \geq 1$. The latter,
parametrized in terms of Generalized Form Factors (GFFs), were  
calculated by both the QCDSF \cite{QCDSF_1} 
and LHPC \cite{LHPC_1}
collaborations for both the unpolarized and polarized 
cases up to $n=3$. At $n\geq 2$, due to the polynomiality property \cite{Die_rev,BelRad}, 
the Mellin moments become dependent on the skewness, $\zeta$.
  
In this contribution we report on the approach used in \cite{AHLT2}; we show our results 
for both proton and neutron DVCS, and we finally present a preliminay result on the 
angular momentum sum rule in the deuteron.

\section{Generalized Parton Distributions from Lattice Moments}

In this contribution we concentrate on the unpolarized scattering
GPDs, $H$, and $E$, from the vector ($\gamma_\mu$)
and tensor ($\sigma_{\mu\nu}$) interactions, respectively. 
We adopt the following set of kinematical variables:
$(\zeta, X, t)$, where $\zeta= Q^2/2(Pq)$
is the longitudinal momentum transfer between the initial 
and final protons ($\zeta \approx x_{Bj}$ in the asymptotic limit,
with Bjorken $x_{Bj} = Q^2/2M\nu$),
$X=(kq)/(Pq)$ is the momentum fraction relative to the initial proton carried by the struck 
parton, $t = -\Delta^2$, is the four-momentum transfer squared. 
$X$ is not directly observable, it appears in the amplitude as an integration variable
\cite{Die_rev,BelRad}.
%%%
The need to deal with a more complicated phase space,  
in addition to the fact that DVCS interferes coherently with the Bethe-Heitler (BH) process,  
are in essence the reasons why it is more challenging to extract GPDs from experiment, 
wherefore guidance from phenomenologically motivated parametrizations becomes important. 

\noindent
We first present a parametrization of $H$ and $E$ in the flavor Non Singlet (NS) sector,
valid in the $X>\zeta$ region, obtained by extending our 
previous zero skewness form \cite{AHLT1}, through proper kinematical shifts:
\begin{equation}
H(X,\zeta,t)  =  G_{M_{X}}^{\lambda}(X,\zeta,t) \, R(X,\zeta,t) 
\label{param1_H}
\end{equation}
%%%%
(a similar form is obtained for $E(X,\zeta,t)$),
where $R(X,\zeta,t)$ is a Regge motivated term  
describing the low $X$ and $t$ behaviors, while
$G_{M_{X}}^{\lambda}(X,\zeta,t)$, was obtained within a spectator model.

\begin{figure}
 %\begin{wrapfigure}{r}{0.5\columnwidth}
\centerline{\includegraphics[width=0.45\columnwidth]%
{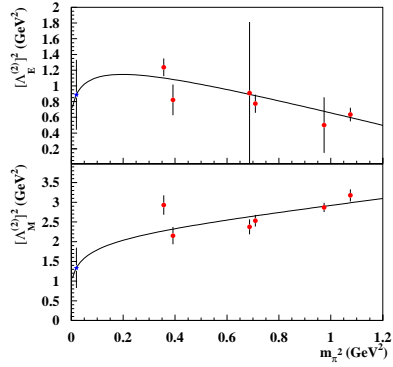}}
\caption{The dipole masses squared for $n=2$, for the isovector magnetic 
(lower panel) and electric (upper panel) contributions obtained by performing fits to the 
lattice results of \protect\cite{QCDSF_1}. 
The value at the physical
pion mass obtained from our fit is also shown (star). (adapted from \cite{AHLT2})}\label{fig1}
%\end{wrapfigure}
  \end{figure}

In order to model the  $X < \zeta$ region, we observe that 
the higher moments of GPDs give 
$\zeta$-dependent constraints, in addition to the ones from the nucleon form factors. 
The $n=1,2,3$ moments of the NS combinations: $H^{u-d} = H^u-H^d$, and $E^{u-d} = E^u-E^d$ 
are available
from lattice QCD \cite{QCDSF_1,LHPC_1}. 
They can be written in terms of the isovector components as: 
\begin{eqnarray}
\label{HuHdn}
H_n^{u-d} \equiv \int \, dX X^{n-1} (H^u - H^d) & = & \frac{\tau (H_M^V)_n + (H_E^V)_n}{1+\tau} \\
\label{EuEdn}
E_n^{u-d} \equiv \int \, dX X^{n-1} (E^u - E^d) & = &  \frac{(E_M^V)_n - (E_E^V)_n}{1+ \tau},
\end{eqnarray}
where the l.h.s. quantities are obtained from the lattice moments 
calculations, whereas $(H_{M(E)}^V)_n$ and $(E_{M(E)}^V)_n$ are amenable to chiral 
extrapolations.
We used lattice calculations for the unpolarized GFFs
obtained by the QCDSF collaboration using two flavors of ${\mathcal O}(a)$-improved 
dynamical fermions for several values of $t$
in the interval 
$0 \lesssim t \lesssim 5$ GeV$^2$, and covering a range of pion mass values, 
$m_\pi \gtrsim 500 \, {\rm MeV}^2$.
Similarly to previous evaluations \cite{LHPC_1} 
the GFFs for both $H$ and $E$, display a dipole type behavior for all three $n$ values,
the value of the dipole mass increasing with $n$. 
We performed an extrapolation by extending 
to the $n=2,3$ moments a simple ansatz proposed in \cite{Ash} for the nucleon form factors 
that: {\it i)} uses the connection between
the dipole mass and the nucleons radius; {\it ii)} introduces a modification of the non analytic 
terms in
the standard chiral extrapolation that suppresses the contribution of chiral loops at large $m_\pi$.
Despite its simplicity, the ansatz seems to reproduce both the lattice results 
trend at large $m_\pi$ 
while satisfying the main physical criteria {\it i)} and {\it ii)}.
Our results for the dipole mass at $n=2$ are shown in Fig.\ref{fig1}.

%%%%%%%%%%%%%%%%%%%
\section{Reconstruction from Bernstein Polynomials}
Similarly to the PDFs case \cite{Yndurain}, 
with a finite number of moments in hand, one can use 
reconstruction methods attaining weighted averages of the 
GPDs, around average ranges of $X$. The weights are provided by 
the complete set of Bernstein polynomials.
The Bernstein polynomials are ideal 
for reproducing the deep inelastic structure functions in that they are zero at the
endpoints, they are normalized to one, and they are peaked in different regions
within the interval $x_{Bj} \in [0,1]$. Because of the latter property the Bernstein
polynomials allow one to emphasize the behavior of the structure function at given
specific regions of $x_{Bj}$, while suppressing the others.  
It was found that $n \geq 8$ moments were necessary to give a fully quantitative
description of the behavior of $F_2(x_{Bj},Q^2)$. 
At present, only $n=3$ Mellin moments are available
from lattice QCD calculations, therefore one cannot reach a similar level of accuracy
as in the inclusive case.%Through method we are able to provide
%an intuitive 
%insight on the behavior of the  function, even if not a point-wise description.  

\begin{figure}%{r}{0.5\columnwidth}
\centerline{\includegraphics[width=0.55\columnwidth]%
{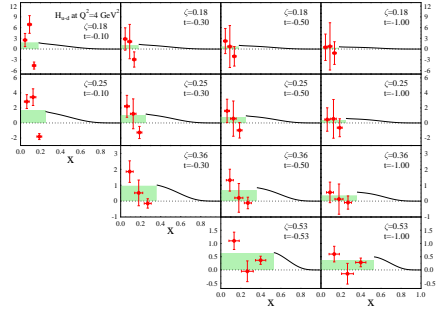}}
\caption{Comparison of $H^{u-d}$ for 
different values of $\zeta = 0.18, 0.25, 0.36, 0.53$, 
and  
$-t \equiv t_{min}= 0.035, 0.073, 0.18, 0.53$ GeV$^2$ 
calculated using the procedure described in the text (adapted from \cite{AHLT2}).}
\label{fig2}
\end{figure}

In Fig.\ref{fig2} we show $H^{u-d}$ reconstructed using the available 
lattice moments. We performed the procedure in the $X<\zeta$ 
region only using:
\begin{equation}
\overline{H}_{k,n}(\zeta,t) = 
\int\limits_0^\zeta H(X,\zeta,t) b_{k,n}(X,\zeta) dX \; \; \; k=0,...n, 
\label{berns1}
\end{equation}
where:
$b_{k,n}(X,\zeta) = X^k \, (\zeta-X)^{n-k}/
\int\limits_0^\zeta X^k \, (\zeta-X)^{n-k} \, dX$, and we used subtracted moments, defined
as: 
\begin{eqnarray}
\label{missing_a}
\left( H_{n} \right)_{X<\zeta}  = H_{n} - \int\limits_\zeta^1 \,  H^I(X,\zeta,t) X^{n} dX  , 
\end{eqnarray}   
\noindent
%\vspace{0.5cm}
where  $H_{n}$ are the Mellin moments, and 
$H^I(X,\zeta,t)$ was obtained from Eq.(\ref{param1_H}). 
For $n=2$, $k=0,1,2$, the reconstruction procedure  
yields \cite{AHLT2}:
%\begin{subequations}
\begin{eqnarray}
\label{berns_zeta}
\overline{H}_{02}( \zeta X_{02})& = & \frac{1}{\zeta^3} \left\{ 3 A_{10}^\zeta \, \zeta ^2 
- 6 A_{20}^\zeta \, \zeta  + 3 
\left[A_{30}^\zeta + \left( -\frac{2\zeta}{2-\zeta} \right)^2 A_{32} \right] \right\},
\nonumber \\
\overline{H}_{12}(\zeta X_{12}) & = & \frac{1}{\zeta^3} \left\{ 6 A_{20}^\zeta \, \zeta - 6 
\left[ A_{30}^\zeta +  \left( -\frac{2\zeta}{2-\zeta} \right)^2 A_{32} \right] \right\},
\nonumber \\   
\overline{H}_{22}(\zeta X_{22}) & = & \frac{1}{\zeta^3} \left\{ 3 A_{30} + 
\left( -\frac{2\zeta}{2-\zeta} \right)^2 A_{32}  \right\} , 
\end{eqnarray}
%\end{subequations}
%%%
where $X_{01}=0.25$, $X_{02}=0.5$, $X_{03}=0.75$, and $A_{10}, A_{20}, A_{30}, A_{32}$ are the
GFFs from Ref.\cite{QCDSF_1}.   

%%%%%%%%%% FIGURE 2

\section{Comparison with Experiment}
\begin{figure}%{r}{0.45\columnwidth}
\centerline{\includegraphics[width=0.45\columnwidth]%
{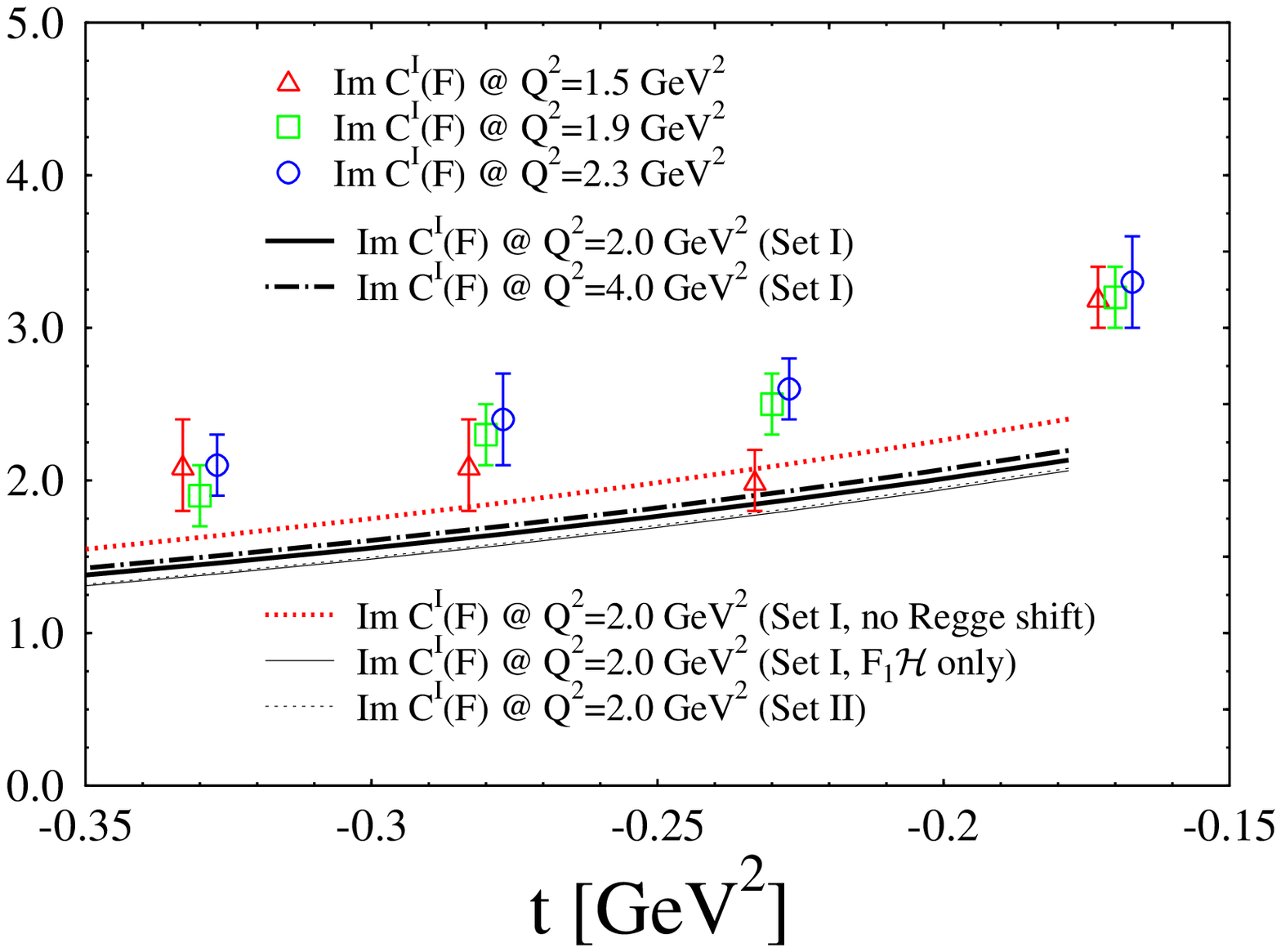}
\includegraphics[width=0.45\columnwidth]%
{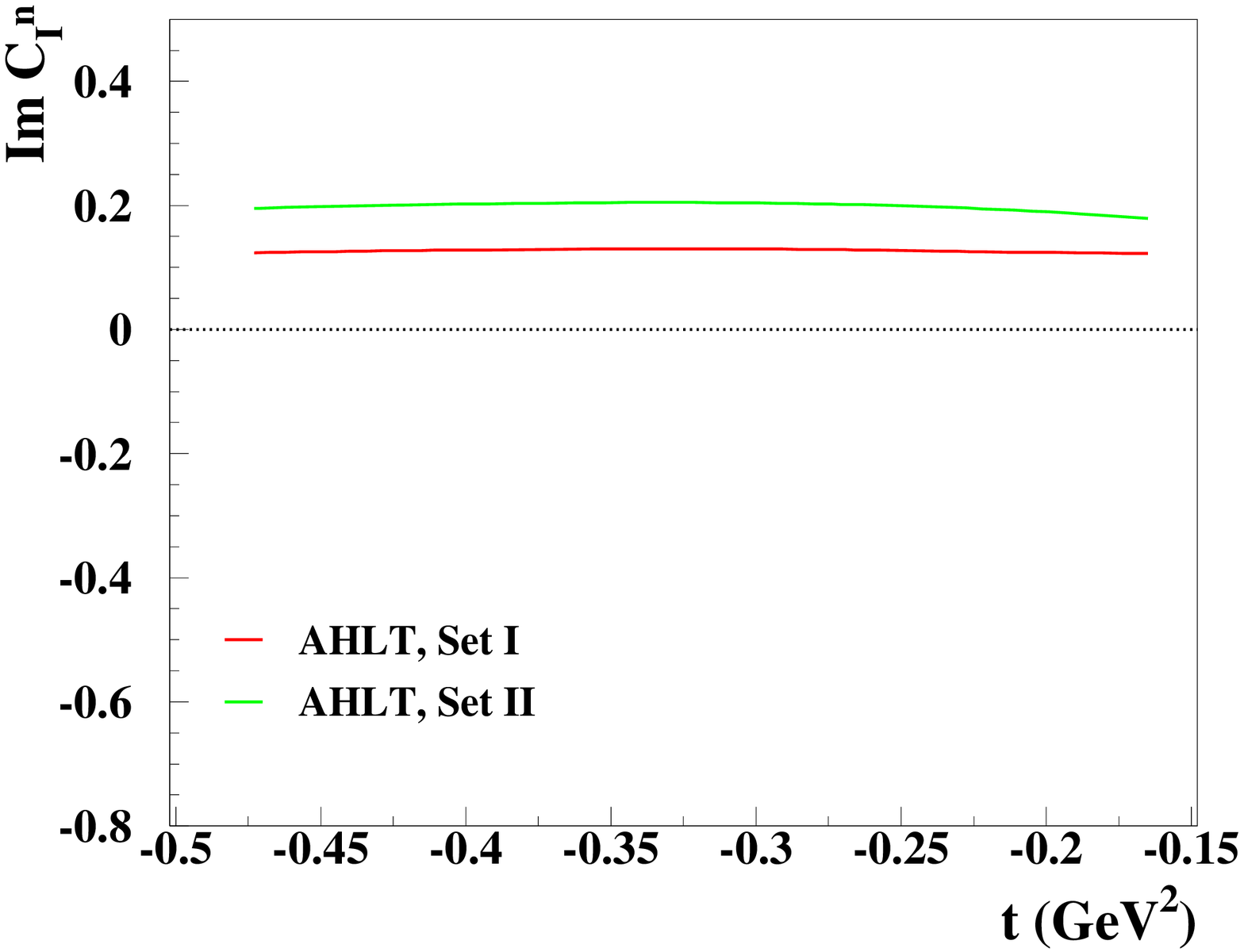}}
\caption{  The imaginary part of 
the Bethe-Heitler BH-DVCS interference term at leading order, $C^I(F)$, (see \protect\cite{Die_rev}).
(upper panel) Proton target. Experimental data from Ref.\cite{hallap}. 
The full curve shows our prediction, including the theoretical error, evolved
to $Q^2=2$ GeV$^2$. The dot-dashed curve, obtained at $Q^2=4$ GeV$^2$, 
shows that the effect of evolution is relatively small.   
The other curves represent variations of the parametrization obtained respectively 
by disregarding the contribution from $E$, by disregarding the 
kinematical shift in the Regge term described in the text, and by using $E$ from 
Set II of Ref.\cite{AHLT1}. 
(lower panel) Neutron target. The two curves 
show our results using variations of the parametrization for $E$ using  
Set I and Set II of Ref.\cite{AHLT1}. The curves are in agreement with the 
experimental data from Ref.\cite{hallan}.
(adapted from Ref.\cite{AHLT2}) }.\label{fig3}
\end{figure}

The unpolarized Compton form factors 
were recently extracted from DVCS on proton and deuteron targets at $Q^2 \approx$ 2 GeV$^2$, 
$\zeta = x_{Bj} = 0.36$, for several values of $t$ in the range $0.15 \leq t \leq 0.35$ GeV$^2$
\cite{hallap,hallan}.   
In Fig.\ref{fig3} and we show the results of our parametrization 
at $\zeta=X$ for the imaginary part of 
the Bethe-Heitler BH-DVCS interference term at leading order, for
proton and neutron, respectively.
The proton results are compared with recent data from Jefferson Lab \cite{hallap}, 
at $x_{Bj}=\zeta=0.36$, and $Q^2 \approx 2$ GeV$^2$, while the neutron
ones are presented at the kinematics of the forthcoming analysis from Ref.\cite{hallan} .

%% figure 3
%% begin figure

\section{Angular Momentum Sum Rule}
We present results derived for the angular momentum for a spin 1 system in terms
of second moments of it's generalized parton distributions (a full derivation
will be shown in \cite{LiuTan}).  

The second moment of the parton distributions
is relevant to the spin
structure of the nucleon. To see this, we first
write down the angular momentum operator in QCD as the sum of
quark and gluon contributions \cite{ji},
\begin{equation}
    \vec{J}_{\rm QCD} = \vec{J}_{q} + \vec{J}_g \ ,
\end{equation}
where
\begin{eqnarray}
     \vec{J}_q &=& \int d^3x ~\vec{x} \times \vec{T}_q \nonumber \\
                 &=& \int d^3x ~\left[ \psi^\dagger
     {\frac{\vec{\Sigma}}{2}}\psi + \psi^\dagger \vec{x}\times (-i\vec{D})
\psi\right]
     \ ,  \nonumber \\
     \vec{J}_g &=& \int d^3x ~\vec{x} \times (\vec{E} \times \vec{B}) \ .
\end{eqnarray}
Here $\vec {T}_q$ and $\vec{E}\times \vec{B}$ are
the quark and gluon momentum densities, respectively. $\vec{\Sigma}$ is
the Dirac spin-matrix and $\vec{D}= \vec{\partial}+
ig\vec{A}$ is the covariant derivative.
By an analogy with the magnetic moment, one can get
the separate quark and gluon contributions to the nucleon spin
if the form factors of the momentum density, or equivalently
the energy-momentum tensor of QCD, are known at zero momentum transfer.
Using Lorentz covariance
and other symmetry principles, one can write down six 
form-factors separately for quark and gluon parts
of the energy-momentum tensor,

\begin{eqnarray}
\langle p' |\theta^{ \mu \nu} |p \rangle=
& - &  \frac{1}{2} \left[P^{\mu}P^{\nu} 
\right](\epsilon'^*  \epsilon)G_{1,2}(t) - \frac{1}{4} \left[P^{\mu}
P^{\nu} \right]
\frac{(\epsilon P)(\epsilon'^* P)}{M^2} G_{2,2}(t) \nonumber \\
& - &  \frac{1}{2} \left[\Delta^{\mu} \Delta^{\nu} - g^{\mu \nu}
{\Delta^2}\right](\epsilon'^*  \epsilon)G_{3,2}(t) \nonumber \\
& - & \frac{1}{4}
\left[\Delta^{\mu}
\Delta^{\nu} - g^{\mu \nu}{\Delta^2}\right]
\frac{(\epsilon P)(\epsilon'^* P)}{M^2} G_{4,2}(t) \nonumber \\
& + & \frac{1}{4} \left[ \left(\epsilon'^{* \mu} (\epsilon P)
 + \epsilon^{\mu} (\epsilon'^* P) \right) P^{\nu} +
\mu \leftrightarrow \nu \right]G_{5,2}(t) \nonumber \\
& + & \left[ \left(\epsilon'^{* \mu} (\epsilon P)
 -  \epsilon^{\mu} (\epsilon'^* P) \right) \Delta^{\nu} +
\mu \leftrightarrow \nu  \right. \nonumber \\  
& + & \left. 2 g^{\mu \nu} (\epsilon P)
(\epsilon'^* P) - \left(\epsilon'^{* \mu} \epsilon^{\nu} +
\epsilon'^{* \nu} \epsilon^{\mu} \right) \Delta^2 \right] G_{6,2}(t)
\end{eqnarray}
where$P^\mu=p^\mu+{p^\mu}'$, $\Delta^\mu = {p^\mu}'-p^\mu$,
and $\epsilon$ is the deuteron polarization.
Taking $\mu=0$, in the Breit frame, the matrix element
of $\vec{J}_q$, gives
\begin{equation}
\langle p'|\int d^{3}x (\vec{x} \times \vec{T}_{q,g}^{0 i})_{z} |p\rangle
= G_{5,2}(0) \int d^{3}x ~ p^{0}
\end{equation}

\begin{equation}
G_{5,2}(0) = 2 {J}^{q}_z 
\end{equation}
Therefore, the generalized form factor, $G_{5,2}(0)$, of the energy momentum tensor
of spin 1 system provides the fraction of the spin is carried by quarks and
gluons.

\section{Conclusions}
In conclusion, we provided a fully quantitative parametrization
of the GPDs in the non-singlet sector, 
valid in the region of Jefferson Lab experiments using both proton and deuteron targets
\cite{hallap,hallan}. 
We also presented a new result for the angular momentum sum rule in the deuteron.
Differently from model calculations, and for the first time to our knowledge, 
our parametrization makes use of experimental data in combination with lattice results. 
Given the paucity of direct experimental measurements of GPDs, and the uncertainties
related to the partonic interpretation of GPDs in the ERBL region,  
our goal is to provide more stringent, model independent  
predictions that will be useful both for model builders, in order 
to understand the dynamics of GPDs, and for the planning of future 
hard exclusive scattering experiments.

\section{Acknowledgments}

We thank Ph. Haegler, P. Kroll and G. Schierholz for useful comments. We are also grateful to 
J. Zanotti for providing us with the recent lattice 
calculations from the QCDSF collaboration.
This work is supported by the U.S. Department
of Energy grant no. DE-FG02-01ER41200 and NSF grant no.0426971. 

% ****************************************************************************
% BIBLIOGRAPHY AREA
% ****************************************************************************

\begin{footnotesize}
% IF YOU DO NOT USE BIBTEX, USE THE FOLLOWING SAMPLE SCHEME FOR THE REFERENCES
% ----------------------------------------------------------------------------

% ----------------------------------------------------------------------------

\end{footnotesize}

% ****************************************************************************
% END OF BIBLIOGRAPHY AREA
% ****************************************************************************

\end{document}